\documentclass[sigconf,authorversion]{acmart}
\usepackage[show]{chato-notes}
\usepackage{graphicx}
\usepackage{epstopdf}
\usepackage{svg}
\usepackage{subcaption}
\graphicspath{ {./images/} }

\definecolor{amaranth}{rgb}{0.9, 0.17, 0.31} 

\newcommand{\sasha}[1]{\textcolor{black}{#1}}
\newcommand{\sfe}[1]{\textcolor{black}{#1}}
\newcommand{\dasha}[1]{\textcolor{black}{#1}}

\setcopyright{acmcopyright}
\copyrightyear{2022}
\acmYear{2022}
\acmDOI{<TBC>}

\acmConference[RecSys CARS '22]{16th ACM Conference on Recommender Systems - Workshop on Context-Aware Recommender Systems}{September 23, 2022}{Seattle, WA, USA}
\acmPrice{15.00}
\acmISBN{<TBC>}

\begin{document}

\title{MTS Kion Implicit Contextualised Sequential Dataset for Movie Recommendation}

\author{Aleksandr Petrov}
\email{a.petrov.1@research.gla.ac.uk}
\affiliation{%
 \institution{University of Glasgow}
 \city{Glasgow}
 \country{United Kingdom}}

\author{Ildar Safilo}
\email{irsafilo@gmail.com}
\affiliation{%
  \institution{MTS, Higher School of Economics}
  \city{Moscow}
  \country{}{Russia}
}

\author{Daria Tikhonovich}
\email{daria.m.tikhonovich@gmail.com}
\affiliation{%
  \institution{MTS}
  \city{Moscow}
  \country{Russia}}

\author{Dmitry Ignatov}
\affiliation{%
  \institution{Higher School of Economics}
  \city{Moscow}
  \country{Russia}}
\email{}

\begin{abstract}

\sasha{We present a new movie and TV show recommendation dataset collected from the real users of MTS Kion video-on-demand platform. In contrast to other popular movie recommendation datasets, such as MovieLens or Netflix, our dataset is based on the implicit interactions registered at the watching time, rather than on explicit ratings. We also provide rich contextual and side information including interactions characteristics (such as temporal information, watch duration and watch percentage), user demographics and  rich movies meta-information. \sfe{In addition, we} describe the MTS Kion Challenge---an online recommender systems challenge that was based on this dataset---and provide an overview of the best performing solutions of the winners. We keep the competition sandbox open, so the researchers are welcome \dasha{to} try their own recommendation algorithms and measure the quality on the private part of the dataset.}
\end{abstract}
\maketitle

\section{Introduction}

Today, recommender systems help us to interact with marketplaces, such as Amazon or eBay, select the most relevant entertainment on streaming platforms, such as Netflix or Spotify, and receive a personalised feed of recommendations in social networks, such as  Twitter or Facebook. \emph{Movie recommendation} is a particular kind of personalisation problem where algorithms select the next videos/movies/series to watch. Modern recommender systems are based on machine learning models (e. g.  Matrix Factorisation-based~\cite{rendle2009bpr} or Deep Learning-based~\cite{sun2019bert4rec, kang2018sasrec, tang2018caser}), and these models need to be trained on user-item interactions data. Some of the models~\cite{rendle2011fast, li2020time} can also use additional contextual and side information, such as item descriptions, user features or features of the user-item interaction (e. g. day of week). Many models also utilise the sequential nature of \dasha{user behaviour}~\cite{sun2019bert4rec, petrov2022effective, kang2018sasrec, petrov2021booking, hidasi2015gru4rec, rendle2010factorizing}. 

Unfortunately, popular movie recommendation datasets, such as MovieLens~\cite{harper2015movielens} and Netflix~\cite{Bennett07netflixprize}, lack this contextual information, and only provide explicit post-view rating feedback. Since meta-information provided by these datasets is limited, they poorly work with cold-start users and items. Another problem of these datasets is that they are somewhat "artificial" - they do not represent actual views of the movies, but rather contain events of rating assignments. Therefore, the sequential nature of video consumption may be broken: rating assignments may be done in a different order compared to the order \dasha{of movie watches}. Often users watch movies without rating them at all. Therefore, a sequence of user's ratings does not always reflect the evolution of the user's interests. As Petrov and Macdonald argue~\cite{petrov2022effective}, with these datasets the task is better described as \emph{the next movie to rate} rather than \emph{the next movie to watch}.  

Hence, we present a new dataset for movie recommendation task that  does not have these limitations\footnote{The dataset available to download on Github: \href{https://github.com/MobileTeleSystems/RecTools/tree/main/datasets/KION}{https://github.com/MobileTeleSystems/RecTools/tree/main/datasets/KION}}. We gathered the data from the users of MTS Kion video streaming platform from 13.03.2021 to 22.08.2021. The dataset includes 5,476,251  interactions of 962,179 users with 15,706 items. Because the dataset is gathered from the real users, it reflects the typical features of a real-world video-on-demand system. \sfe{We only perform a} minimal preprocessing of the data to preserve user's privacy\footnote{The only reprocessing we do is anonymisation and  adding a small amount of noise in order to preserve the privacy of Kion's users}. For example, in this dataset one can observe strong 
\sfe{periodical} patterns and strong \sfe{time-depended trends}. The dataset includes both cold and warm users and items, respectively as it often happens in the real-world applications. We provide more details about these features of the dataset in Section~\ref{sec:exploratory_analysis}

With the rich meta-information, the dataset enables research  beyond accuracy metrics. For example, the meta-information enables measuring such metrics as diversity, novelty, and serendipity~\cite{shani2022eval}. In contrast to other popular movie datasets, it includes user demographic features,  such as age, sex, and income level,  and also contains rich movie features, such as descriptions, directors, actors and countries of production. This meta-data may help to analyse the deep relationships between the users and the items. For example, we have discovered that the proportion of female users among the movie watchers is a very strong feature for the recommendation models. 

We used this dataset for the MTS Kion recommender systems challenge that we ran in November 2021 - January 2022. During the time of the competition, 46 contestants submitted 844 solutions. We provide an overview of the winning recommendation models in  Section~\ref{sec:competition}. The importance of the meta-information, presented in the dataset, was confirmed during the Kion challenge by all of the winning solutions.

Overall, the dataset enables researchers to build more complex models and to perform more deep and interesting analysis compared to the other existing datasets. 

In short, our contributions may be described as follows: 
\begin{enumerate}
    \item We provide researchers with a new implicit contextualised sequential dataset for movies recommendation. 
    \item We analyse the salient features of the dataset. 
    \item We review the recommendation models, which were built by the winners of the recommender systems challenge that we ran using the dataset. 
    \item We release a public sandbox, where researchers can benchmark the quality of their own recommendation models on the private part of the dataset. 
\end{enumerate}

In the  rest of the paper we provide more details about the dataset and the Kion challenge. In particular, Section~\ref{sec:exploration} provides an overview of the salient characteristics of the dataset, including deeper analysis of the data. Section~\ref{sec:competition} reviews the MTS Kion challenge and describes the selected contestant's solutions. Section~\ref{sec:conclusion} contains the final remarks. 

\section{Kion Dataset} \label{sec:exploration}

\subsection{Kion Video-On-Demand platform}
In order to help researchers to better understand the nature of our dataset, we  briefly review the Kion platform that provided us with the data.

Kion is one of the biggest video streaming platforms in the Eastern Europe with more than 3,5 millions subscribers. At the same time,  Kion is a relatively new platform founded by the MTS holding in 2021; therefore, the dataset includes the patterns of rapid growth from the foundation of the service to 1 million users. We now provide more details about the dataset.  

\subsection{Dataset Description}

Kion dataset is an implicit contextualised sequential dataset with information about user-item interactions. It provides interactions over a 5 months period starting on 13.03.2021 and ending on 22.08.2021. The dataset contains 5,476,251 interactions of 962,179 users with 15,706 items. Each interaction is described with the date, total watch duration and percent of the content watched by the user. In case of multiple interactions with the same item we only include the last interaction in the dataset, however total duration and percentage is aggregated over all watches. 

We also provide side information for both users and items. Users are described with demographics, such as age, sex, income and "has kids" flag. Items have such features as content type (movies or series), title, title in the original language, release year, genres, countries, age rating, studios, directors, actors, keywords and description.

\subsection{Exploratory Analysis} \label{sec:exploratory_analysis}

\begin{figure*}[tb]
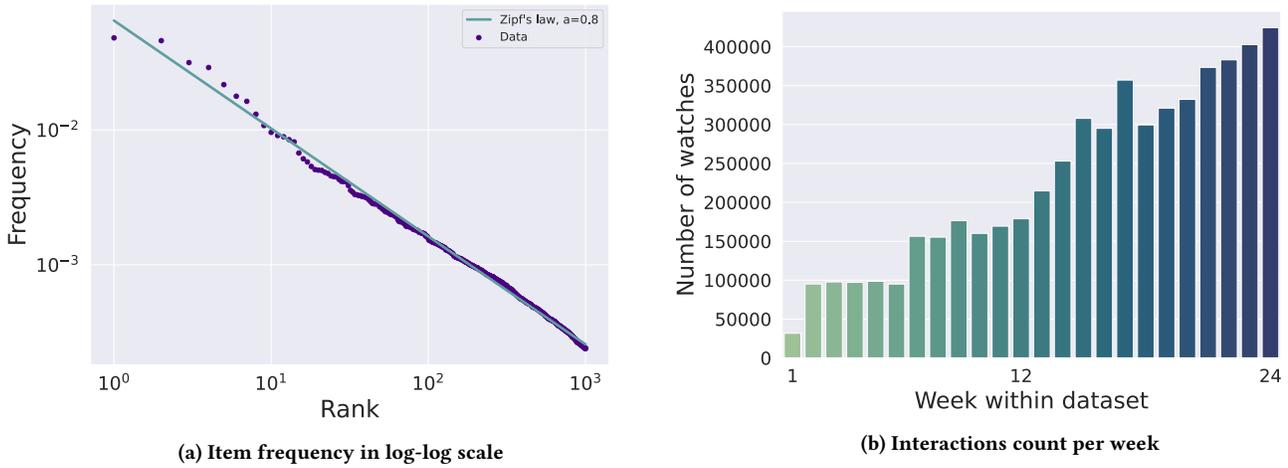

\begin{subfigure}{.5\textwidth}
    \includesvg[width=\linewidth]{zipf_law.svg}
    \caption{Item frequency in log-log scale}
    \label{fig:interactions_per_item}
\end{subfigure}%
\begin{subfigure}{.5\textwidth}
    \includesvg[width=\linewidth]{dist_by_week.svg}
    \caption{Interactions count per week}
    \label{fig:interactions_per_week}
\end{subfigure}%
\caption{Interactions distribution}
\label{fig:interactions_distribution}
\end{figure*}

\begin{figure*}[tb]
    \includesvg[width=\linewidth]{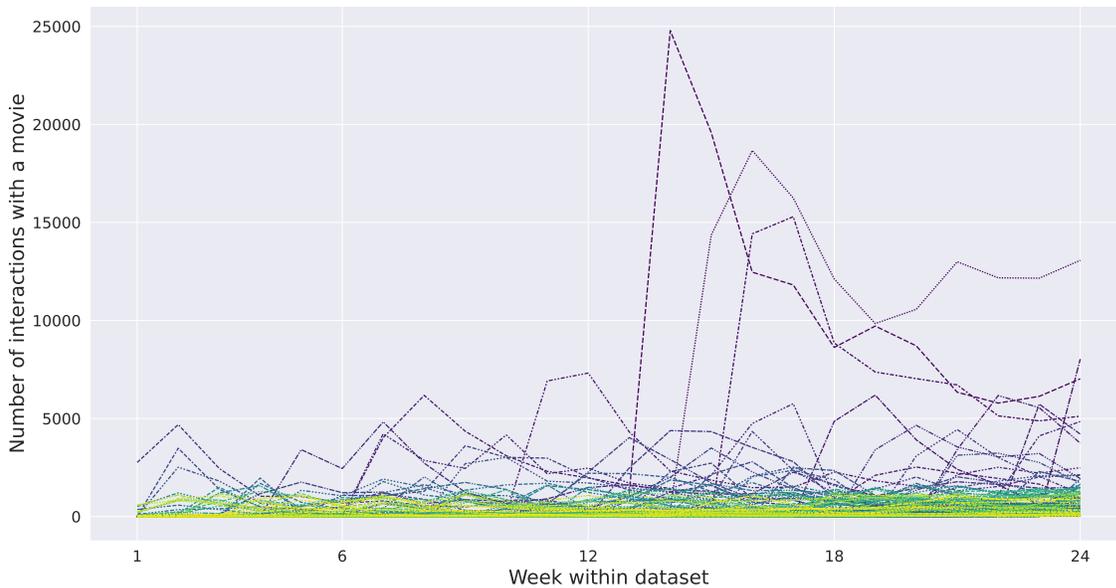}
    \caption{Top 100 movies: interactions count per week}
    \label{fig:top100_int_per_week}
\end{figure*}

Datasets with movie interactions usually have power-law distribution with a few items dominating in user preferences~\cite{steck2021deep}. Figure ~\ref{fig:interactions_per_item} illustrates this fact for Kion dataset, showing item frequencies in a log-log scale. As the figure shows, the data can be approximated quite well with Zipf's law distribution with degree parameter $\alpha=0.8$.
It is also important to emphasise clear time trends of items popularity. Figure~\ref{fig:interactions_per_week} shows the growing number of interactions each week in Kion dataset. Figure~\ref{fig:top100_int_per_week} illustrates individual jumps of popularity of top 100 movies (excluding TV shows). Not only some of the items have huge excess weight over the majority of the catalogue, but also top items popularity can vary greatly during a short period of two or three weeks.
These particular qualities of popularity distribution in time and within items may have notable consequences for building a correct validation scheme and successfully solving the movie recommendation problem on such data.

\begin{figure*}[ht]
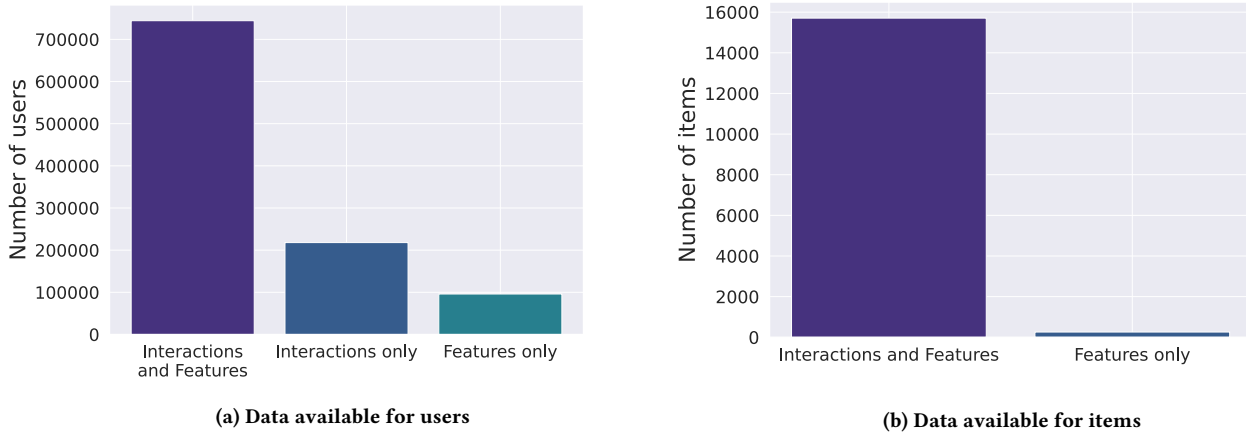

\begin{subfigure}{.5\textwidth}
  \includesvg[width=.95\linewidth]{users_data.svg}
  \caption{Data available for users}
  \label{fig:user_features}
\end{subfigure}%
\begin{subfigure}{.5\textwidth}
  \includesvg[width=.95\linewidth]{items_data.svg}
  \caption{Data available for items}
  \label{fig:items_features}
\end{subfigure}%
\caption{Features and interactions data availability}
\label{fig:data_availability}
\end{figure*}

User and item features are another important aspect for recommender systems. The vast majority of the items in the dataset have both features and interactions, while 30 percent of the users have only one of the above. Figure ~\ref{fig:data_availability} illustrates data availability for both users and items.

Overall, Kion dataset reflects specific challenges of the real world recommender systems, including:
(i) implicit feedback, (ii) Zipf's distribution of items popularity, (iii) time-depended spikes of popularity of individual items, (iv) cold-start problem for both users and items, (v) diverse but incomplete meta-information.

\subsection{Comparison with Existing Movie Recommendation Datasets}\label{ssec:existing}

In this section we compare the Kion dataset with other movie recommendation datasets available prior to our work.  In particular, we compare the Kion dataset with MovieLens-25M,  the largest dataset of the MovieLens family~\cite{harper2015movielens}, and with the Netflix dataset, which was released for the famous Netflix Prize challenge~\cite{Bennett07netflixprize}. 

We first compare the quantitative characteristics of the datasets, which are presented in Table~\ref{table:datasets_comparsion_quantitative}. As we can see from the table, the number of users in the Kion dataset is by far larger compared to both Netflix (+100\%) and MovieLens (+493\%). However, at the same time the number of interactions in Kion is smaller compared to the other two datasets. These two facts lead to much shorter average interaction sequences: in Kion average sequence length is just 5.69, compared to 153.80 in MovieLens and 209.25 in Netflix. We argue that these shorter sequences and large numbers of cold-start users are common in the real-world recommender systems, especially during the rapid growth phase. 

We now compare the qualitative characteristics of the datasets, which are presented in Table~\ref{table:datasets_comparsion_qualitative}. As we can see, Netflix and MovieLens have quite similar characteristics: they are both explicit ratings datasets that are collected some time after movie watching events. They both provide only limited contextual information about movies and do not provide users data at all.  In contrast, Kion is an implicit dataset, with events registered at the watching time. It includes rich meta-information for both users and items. It also includes additional information about events themselves, such as duration and percentage of the view. 

Overall, as we can see from both Tables ~\ref{table:datasets_comparsion_quantitative} and~\ref{table:datasets_comparsion_qualitative} the Kion dataset has very different qualitative and quantitative features compared to the existing datasets and may facilitate new kinds of research in movies recommendation field that were not possible with MovieLens or Netflix.

\begin{table*}[tb]
\caption{Quantitative characteristics of movie datasets}\label{table:datasets_comparsion_quantitative}
\begin{tabular}{llllll}
\toprule
\textbf{Dataset Name} & \textbf{Users} &\textbf{Items}& \textbf{Interactions} & \textbf{Avg. Sequence Length} & \textbf{Sparsity} \\
\midrule
Netflix & 480,189 & 17,770 & 100,480,507 & 209.25 & 98.82\% \\
Movielens-25M & 162,541 & 59,047 & 25,000,095 & 153.80 & 99.73\% \\
\textbf{Kion} & \textbf{962,179} & \textbf{15,706} & \textbf{5,476,251} & \textbf{5.69} & \textbf{99.9\%} \\
\bottomrule
\end{tabular}
\end{table*}

\begin{table*}[tb]
\caption{Qualitative characteristics of movie datasets}\label{table:datasets_comparsion_qualitative}
\begin{tabular}{|l|l|l|l|}
\hline
\textbf{Dataset Name} & \textbf{Netflix} & \textbf{Movielens-25M} & \textbf{Kion} \\ \hline
\textbf{Type} & Explicit (Ratings) & Explicit (Rating) & Implicit (Interactions) \\ \hline
\textbf{Interaction registration time} & After watching & After watching & At watching \\ \hline
\textbf{Interaction features} & Date, Rating & Date, Rating & Date, Duration, Watched Percent \\ \hline
\textbf{User features} & None & None & Age, Income, Kids \\ \hline
\textbf{Item features} & Release Year, Title & \begin{tabular}[c]{@{}l@{}}Release Year,\\ Title, \\ Genres, Tags\end{tabular} & \begin{tabular}[c]{@{}l@{}}Content Type, Title, \\ Original Title, Release Year, \\ Genres, Countries, For Kids,\\  Age Rating, Studios, Directors, \\ Actors, Description, Keyword\end{tabular} \\ \hline
\end{tabular}
\end{table*}

\subsection{English Translation of the Meta-data}
Most of the users of Kion service are Russian-speaking. Therefore, it stores the movies meta-data, such as titles, descriptions, actors and directors in Russian language. In order to make the dataset more accessible for the researchers around the world, we translate this meta-information to English using Facebook FAIR’s WMT19 Ru->En machine translation model~\cite{ng2019facebook} within Hugging Face Transformers~\cite{wolf2019huggingface} machine translation pipeline. For the names we also use transliteration available in the python transliterate package\footnote{\href{https://pypi.org/project/transliterate/}{https://pypi.org/project/transliterate/}}. Despite the imperfections of the automated translation, we find that the results are acceptable and can be used as the auxiliary features for the recommender models. We publish both original Russian and translated English versions of the meta-data. 

\section{Kion Recommender Systems Competition}
\label{sec:competition}
\subsection{Competition Details}
We used the dataset for an online competition, which we conducted in November 2021 - December 2022.  
We \sasha{hosted the} competition on the ODS.ai platform \footnote{\href{https://ods.ai/competitions/competition-recsys-21}{https://ods.ai/competitions/competition-recsys-21}}. 
The main goal of the competition was to predict movies that the users watched within one week following the period covered by the dataset. We keep these interactions in the \emph{private} part of the dataset, which was not available to the participants. 

\subsubsection{Competition Metric}
\begin{figure}[tb]
    \includesvg[width=\linewidth]{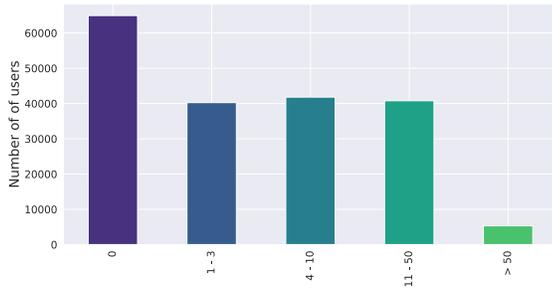}
    \caption{Interactions count per user in test dataset}
    \label{fig:test_users_stats}
\end{figure}

As the main performance metric we used \emph{Mean Average Precision} at cutoff $K$ ($MAP@K$), which we measured on the private dataset. One third of the users from the private dataset can be considered "cold" (i. e., these users do not have interactions during the train period). Figure~\ref{fig:test_users_stats} shows the distribution of users by the number of interactions in the public dataset.

Mean Average Precision is a precision-based metric. Consider a list of recommendations $\{i_1, i_2, ..., i_n\}$ for a user with ground truth binary relevance labels $\{r_1, r_2, ..., r_n\}$. Then, \emph{Precision} at cutoff $K$ ($P@K$) is defined as

\begin{align}
     P@K  = \frac{\sum_{i=1}^{k}{r_i}}{K}
\end{align}

\emph{Average Precision} at cutoff $K$ ($AP@K$) for the user is defined by averaging Precisions and normalising them: 

\begin{align}
    AP@K(user) = \frac{1}{C_{user}} \sum_{i = 1}^{K}{P@i \cdot r_i}
\end{align}

where $C_{user}$ is the total count of relevant items for the user in the private dataset. To get \emph{Mean Average Precision} at cutoff $K$ we average $AP@K$ metric over the dataset. 

\begin{align}
    MAP@K = \sum_{i = 1}^{N}{AP@K(user_i)}
\end{align}
where $N$ is the number of users in the private dataset. 
For the competition we chose cutoff 10, so the main metric of the competition was $MAP@10$. 
\subsection{Competition sandbox}
We do not release the private dataset publicly, however we keep the competition sandbox open\footnote{\href{https://ods.ai/competitions/competition-recsys-21/leaderboard/public\_sandbox}{https://ods.ai/competitions/competition-recsys-21/leaderboard/public\_sandbox}}, so other researchers are welcome to make a submission and measure their $MAP@10$ result measured on the private part of the dataset. 

To make a submission, one needs to use \emph{sample\_submission.txt} file, and replace the sample item ids with the ids of the recommended items according to a recommendation model. 

In the next section we provide an overview of some of the notable solutions, provided by the contestants. 

\subsection{Notable Solutions}

Table~\ref{table:results} contains results of the top 5 best participants measured on the private dataset. All of the winners used a two-stage approach, generating candidates for recommendations with a single or multiple base models at the first stage and rearranging them with Gradient Boosting Trees on the second stage, using outputs from the first-stage-models along with additionally engineered features.

\begin{table*}[tb]

\caption{Results of the best-performing participants.}\label{table:results}

\begin{tabular}{llll}
\toprule
\textbf{Position} & \textbf{Name}                         & \textbf{MAP@10} & \textbf{Solution Type} \\
\midrule
1                 & Oleg Lashinin & 0.1221    & Neural and Non-Neural ensemble \\
2                 & Aleksandr Petrov                      & 0.1214    & Neural and Non-Neural ensemble \\
3                 & Stepan Zimin                          & 0.1213     & Non-Neural ensemble     \\
4                 & Daria Tikhonovich                     & 0.1148     & Gradient Boosting Trees    \\
5                 & Olga                    & 0.1135     & Gradient Boosting Trees    \\
\midrule
\multicolumn{2}{l}{\textbf{Popularity Baseline}}          & \textbf{0.0910} \\
\bottomrule
\end{tabular}
\end{table*}

\subsubsection{Neural Ensembles}
Participants Oleg Lashinin and Aleksandr Petrov, who achieved the first and the second positions in the final leaderbord, used 2-stage ensembles with multiple recommender models at the first stage. The ensembles included a mixture of strong deep-learning based models -- including BERT4Rec~\cite{sun2019bert4rec}, SASRec~\cite{kang2018sasrec} and Caser~\cite{tang2018caser} -- and traditional recommendation approaches, such as Bayesian Personalised Rank~\cite{rendle2009bpr}. Oleg used implementations of the neural models from RecBole library~\cite{zhao2021recbole} and implementations of some of the models published by Dacrema et al.~\cite{dacrema2019we}, whereas Aleksandr used the implementations described in replicability paper~\cite{petrov2022replicability}

To ensemble the first-stage models,  the participants used Gradient Boosting Trees from the LightGBM library~\cite{ke2017lightgbm}. Oleg used a classification training objective, and Aleksandr used the LambdaRank~\cite{burges2010ranknet} ranking objective. 

Both winning participants supplemented the outcomes of the first-stage models with the manually engineered features, such as time since last user's interaction or proportion of men among movie viewers. 

\subsubsection{Non-neural solutions}
Participant Stepan Zimin followed the same approach as the top 2 winners, building an ensemble of the first-stage models, but didn't include any deep learning models. Their solution was based on the output of the five different baselines, including a hybrid matrix factorisation model LightFM~\cite{kula2015metadata}, which utilises user and item meta-data. They enriched the outcome of the models with diverse interactions statistics of both users and items and also included item genres while building features for the ensemble. They used Gradient Boosting Trees from the CatBoost~\cite{dorogush2018catboost} library with the YetiRank~\cite{gulin2011winning} ranking objective for the final predictions and achieved MAP@10 score which was almost identical to the top 2 solutions.

Participants Daria Tikhonovich and Olga, who took the fourth and the fifth winning places in the competition, achieved competitive MAP@10 scores with only one model at the first stage. Daria used an item-item nearest neighbour model (Item KNN) on user-item matrix~\cite{deshpande2004item} and Olga used a simple popularity model. Both participants used Gradient Boosting Trees from the CatBoost library with a classification training objective at the second stage. Daria added manually generated features based on items popularity trends (e. g. number of interactions during last weeks, popularity trend slope from the last week and distance in days to the 0.95 quantile of the item interactions dates distribution). Olga utilised items from user history as features. Both participants also added demographic characteristics of users.

Overall, the top 5 winning solutions of the competition on the Kion dataset utilised a wide range of approaches, including sequential recommendation models (BERT4Rec, SASRec, Caser), contextualised models (LightFM) and classical recommendation approaches (Matrix Factorisation, Item KNN). All of the winners enriched model outputs with diverse features exploiting user and item meta-data as well as interactions statistics. This broad variety of successful approaches to the recommendation task of the competition shows the value of the dataset for recommender systems researchers and the comprehensive nature of its data.

\section{Conclusion}\label{sec:conclusion}

In summary, we offer a novel movies recommendation dataset from the MTS Kion streaming platform.  We  compared it with  MovieLens-25M and Netflix datasets and concluded that our dataset better represents real-world challenges: (i) implicit
feedback, (ii) Zipf’s distribution of items popularity, (iii) time-depended spikes of popularity of individual items, (iv)
cold-start problem for both users and items, (v) diverse but incomplete meta-data. We have also described the best solutions of the MTS Kion Challenge which was based on our dataset. We hope that our paper and its online appendices will spark further research in the field of contextualised and sequential recommender systems.

\section{Acknowledgements}
We would would like to acknowledge Kion challenge participants Oleg Lashinin,  Stepan Zimin, and Olga for providing descriptions of their Kion Challenge solutions, MTS Holding for providing the Kion dataset, ODS.ai international platform for hosting the competition. 

\balance
\bibliographystyle{ACM-Reference-Format}
\bibliography{references}

\end{document}